\documentclass{article}
\usepackage{arxiv}
\usepackage[utf8]{inputenc} 
\usepackage[T1]{fontenc}    
\usepackage{hyperref}       
\usepackage{url}            
\usepackage{booktabs}       
\usepackage{amsfonts}       
\usepackage{nicefrac}       
\usepackage{microtype}      
\usepackage{multicol}
\usepackage{graphicx}
\usepackage{adjustbox}
\usepackage{placeins}
\usepackage{algorithm} 
\usepackage{algpseudocode} 
\graphicspath{ {./images/} }
\usepackage{listings, xcolor}

\definecolor{verylightgray}{rgb}{.97,.97,.97}

\lstdefinelanguage{Solidity}{
	keywords=[1]{anonymous, assembly, assert, balance, break, call, callcode, case, catch, class, constant, continue, constructor, contract, debugger, default, delegatecall, delete, do, else, emit, event, experimental, export, external, false, finally, for, function, gas, if, implements, import, in, indexed, instanceof, interface, internal, is, length, library, log0, log1, log2, log3, log4, memory, modifier, new, payable, pragma, private, protected, public, pure, push, require, return, returns, revert, selfdestruct, send, solidity, storage, struct, suicide, super, switch, then, this, throw, transfer, true, try, typeof, using, value, view, while, with, addmod, ecrecover, keccak256, mulmod, ripemd160, sha256, sha3}, 
	keywordstyle=[1]\color{blue}\bfseries,
	keywords=[2]{address, bool, byte, bytes, bytes1, bytes2, bytes3, bytes4, bytes5, bytes6, bytes7, bytes8, bytes9, bytes10, bytes11, bytes12, bytes13, bytes14, bytes15, bytes16, bytes17, bytes18, bytes19, bytes20, bytes21, bytes22, bytes23, bytes24, bytes25, bytes26, bytes27, bytes28, bytes29, bytes30, bytes31, bytes32, enum, int, int8, int16, int24, int32, int40, int48, int56, int64, int72, int80, int88, int96, int104, int112, int120, int128, int136, int144, int152, int160, int168, int176, int184, int192, int200, int208, int216, int224, int232, int240, int248, int256, mapping, string, uint, uint8, uint16, uint24, uint32, uint40, uint48, uint56, uint64, uint72, uint80, uint88, uint96, uint104, uint112, uint120, uint128, uint136, uint144, uint152, uint160, uint168, uint176, uint184, uint192, uint200, uint208, uint216, uint224, uint232, uint240, uint248, uint256, var, void, ether, finney, szabo, wei, days, hours, minutes, seconds, weeks, years},	
	keywordstyle=[2]\color{teal}\bfseries,
	keywords=[3]{block, blockhash, coinbase, difficulty, gaslimit, number, timestamp, msg, data, gas, sender, sig, value, now, tx, gasprice, origin},	
	keywordstyle=[3]\color{violet}\bfseries,
	identifierstyle=\color{black},
	sensitive=false,
	comment=[l]{//},
	morecomment=[s]{/*}{*/},
	commentstyle=\color{gray}\ttfamily,
	stringstyle=\color{red}\ttfamily,
	morestring=[b]',
	morestring=[b]"
}

\title{Blockchain-Based Decentralized Knowledge Marketplace Using Active Inference}

\author{
 Shashank Joshi \\
 Department of Computer Science and Engineering\\
 SRM Institute Of Science And Technology\\
 Kattankulathur, Tamil Nadu – 603203, \\
 \texttt{sj8559@srmist.edu.in} \\
   \And
Arhan Choudhury \\
 Department of Computer Science and Engineering\\
 SRM Institute Of Science And Technology\\
 Kattankulathur, Tamil Nadu – 603203, \\
 \texttt{ac8365@srmist.edu.in} \\
\And
}

\begin{document}
\maketitle

\begin{abstract}
\textbf{A knowledge market can be described as a type of market where there is a consistent supply of data to satisfy the demand for information and is responsible for the mapping of potential problem solvers with the entities which need these solutions. It is possible to define them as value-exchange systems in which the dynamic features of the creation and exchange of intellectual assets serve as the fundamental drivers of the frequency, nature, and outcomes of interactions among various stakeholders. Furthermore, the provision of financial backing for research is an essential component in the process of developing a knowledge market that is capable of enduring over time, and it is also an essential driver of the progression of scientific investigation. This paper underlines flaws associated with the conventional knowledge-based market, including but not limited to excessive financing concentration, ineffective information exchange, a lack of security, mapping of entities, etc. The authors present a decentralized framework for the knowledge marketplace incorporating technologies such as blockchain, active inference, zero-knowledge proof, etc. The proposed decentralized framework provides not only an efficient mapping mechanism to map entities in the marketplace but also a more secure and controlled way to share knowledge and services among various stakeholders.
}
\end{abstract}


\section{Introduction}
To be successful in today's extremely competitive and knowledge-based economy, businesses need to be able to handle huge amounts of information and services effectively. This kind of information and service is extremely sophisticated, not to mention that it is incredibly diverse, and it is expanding at an ever-increasing rate. Over the course of time, different knowledge management strategies have emerged \cite{1}, and It is apparent that the preponderance of the studies and developments have been concentrated on the management of knowledge within organizations, which is where a large number of issues have been resolved \cite{2}. Nevertheless, this opens up a window of opportunity for organizations or individuals to economically utilize their knowledge assets \cite{3}, and the phenomena of utilizing the crowd's knowledge assets against commercialization give rise to the concept of knowledge markets. Nevertheless, the existing knowledge monetizing ecosystem is inefficient and thus undermines the productivity of the whole knowledge-sharing sector due to its inherent structural and operational intricacies.\\
To begin with, research funding is a crucial facet of a sustaining knowledge marketplace, and the structure of the current knowledge market is inherently characterized by the disparity in research spending \cite{4}. Furthermore, one of the main causes of this discrepancy is an improper mapping of the parties who would be interested in a research consignment (which usually culminates in the interchange and commercialization of knowledge assets). In addition, the fact that prominent players are involved in this sector is another factor that adds to the problem of excessive funding concentration \cite{5}. Next, the conventional knowledge market framework is also cluttered with a plethora of intricacies, which, in turn, leads to inefficient information flow between the stakeholders. Lastly, other problems associated with the conventional knowledge trade include a lack of information security, privacy, access control, ownership transfer, and transparency.\\
In order to resolve the aforementioned issues and to streamline the exchange of knowledge assets, this paper proposes a novel decentralized framework in order to construct a trustworthy and functional knowledge marketplace, which employs active inference-based free energy analysis for the efficient mapping of the stakeholders for a research contract along with blockchain's innate qualities including but not limited to immutability, secure, distributed, fraud resilience etc. which based on its cutting-edge data cryptographic techniques, assures both the security of the data as well as the user's identification. To avoid the issues of a centralized framework in which all information and knowledge is held by a limited number of parties, the proposed framework makes use of current advancements in computing technology to establish a flexible and distributed network. To facilitate each party's contribution, the framework allows for not only the sharing of information but also the trading of expertise and services.\\
The paper addresses four foundational facets of the knowledge marketplace. First, it verifies and assesses whether or not the researchers and the investors are a good fit for one another. Second, the proposed framework guarantees equitable monetization, meaning that both the investor and the knowledge seller should benefit from the knowledge monetization, and the knowledge buyer should obtain the knowledge after payment has been received. Third, ensure that the information is not divulged to any other parties until the process of knowledge monetization has been fully completed, thus ensuring the confidentiality of intellectual assets. Finally, the proposed architecture also streamlines all the interactions and processes among various stakeholders of the network.\\ 
The remaining sections of this paper are structured as follows: Section 2 provides some context or preliminary information on this work. Next, Section 3 provides a summary of the current attempts to digitize the knowledge economy, including a review of the relevant research and associated works. The working of the knowledge marketplace is outlined in section 4. Section 5 of the paper then describes the conventional knowledge exchange system along with its flaws. Then, Section 6 elaborates on the proposed framework for the knowledge marketplace and its potential outcome on the knowledge monetization sector. Finally, The paper is concluded in section 7, along with some recommendations for further research.
\section{Background}
\subsection{Blockchain}
A decentralized distributed ledger that can only be appended to and has a synchronization mechanism constitutes a blockchain network. It logs every transaction and verifies the origin of any assets involved in those transactions in chronological order using cryptographic links between each block. Blockchain technology, in contrast to centralized apps found on the internet, does not have a single controlling authority. The distributed ledger system (blockchain) consists of a network of nodes, each of which copies the ledger. It is seen in a public capacity by every user that is connected to the network \cite{6}. Because only the wallet address may be matched to a transaction, all of the transactions are open to public scrutiny while maintaining a veneer of anonymity \cite{7}. The first practical implementation of blockchain technology was the "cryptocurrency" Bitcoin \cite{8}, which is also known as a "decentralized digital currency" that is verified by encryption. After that, a plethora of brand-new currencies, each with their own set of distinctive characteristics, such as Ethereum \cite{9}, emerged. The basic architecture of a blockchain network is illustrated in Figure 1.
\begin{figure} 
    \centering
    \begin{center}
    \includegraphics[scale=1]{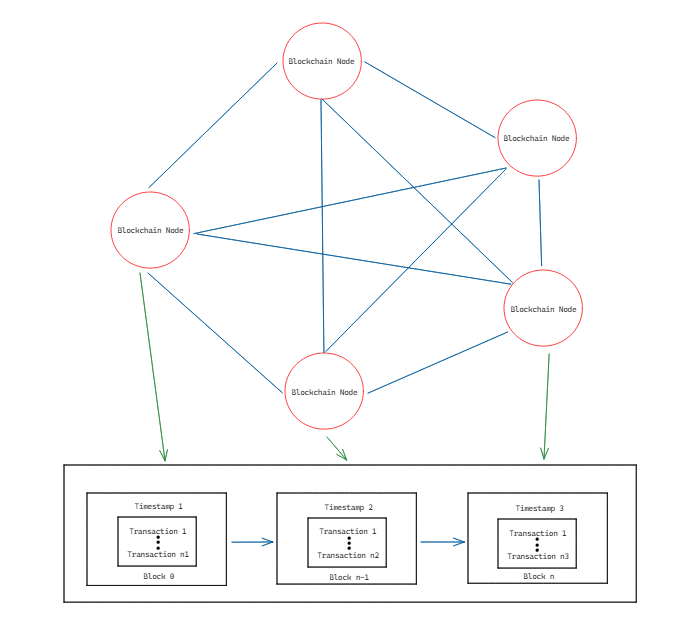}
    \end{center}
    \caption{Architecture Of Blockchain Network}
\end{figure}
\subsection{Active Inference}
Active inference is a theory that describes perception, planning, and action in terms of probabilistic inference. It is a way of describing sentient activity \cite{10}. Active inference, which theoretical neuroscientist Karl Friston created during years of ground-breaking research, offers an integrated viewpoint on the brain, cognition, and behaviour that is increasingly employed in various fields, including neuroscience, psychology, and philosophy. Active inference translates observation into action. Active inference's theory, applications, and cognitive domains simulate various complex systems.\\
By framing behaviour and the brain in terms of a single drive to reduce free energy, active inference is a "first principles" approach to understanding behaviour and the brain. In order to comprehend how the brain functions, there is a strong emphasis on the implications of the free energy principle. In order to contextualize active inference within the most recent theories of cognition, it is first introduced theoretically and formally. It then offers concrete illustrations of computer models that employ active inference to describe many aspects of cognition, including perception, attention, memory, and planning.\\
According to Active Inference, the brain approximates the best possible probabilistic (Bayesian) belief updating in the perceptual, cognitive, and motor domains. More specifically, Active Inference holds that the brain has a "generative" internal model of the world that can mimic the sensory input it should experience if its model of the world is accurate. The differences between predicted and observed sensations can be used to update the model by comparing the simulated (predicted) sensory data against actual observations.\\
This updating corresponds to inference (perception) over short durations (like a single observation), whereas it corresponds to learning over longer timescales (i.e., updating expectations about what will be observed later) \cite{11}. Another way to say it is that learning optimizes ideas about the connections between the variables that make up the universe, whereas perception optimizes beliefs about the world as it is now. These procedures can be thought of as ensuring that the generative model, which the brain generates due to its recognition processes, accurately represents the outside world \cite{12}.

\subsection{Decentralized Autonomous Organization (DAO)}
A decentralized autonomous organization, also known as a DAO, is a blockchain-based system that enables people to coordinate and self-govern themselves using a set of self-executing rules that are deployed on a public blockchain \cite{13}. DAOs have decentralized governance and are independent of central control. DAOs work in a manner analogous to organizations that mediate interactions between a group of individuals, often an open community that recruits new members. Members of certain decentralized autonomous organizations (DAOs) function analogously to stockholders in traditional businesses in that they possess a particular token that enables participation in the DAO.\\
DAOs are considered to be autonomous because, unless their code specifies expressly otherwise, they operate independently from the developers who created them. The organization's activities, which are governed by its members (humans), are compliant with the standards established in its code. Additionally, due to the fact that they are deployed on a public blockchain, they are resistant to censorship. This is because there is no centralized authority that has the ability to shut down the DAO and the service that it provides. DAOs will therefore continue to function, for example, by providing services, buying and selling resources, or hiring people, so long as there are members willing to execute their code \cite{14}. This might mean that DAOs continue to give services, buy and sell resources, or hire people.

\subsection{Smart Contract}
Smart contracts facilitate, carry out, and enforce agreements between unreliable parties \cite{15}. This is because smart contracts do not require the support of a trustworthy third party. They function atop the blockchain network and make it possible to digitize contracts that were previously only available on paper. Solidity, which is a Turing-complete language, is the language that is used for the generation of smart contracts on the Ethereum network. They offer many of the same functionalities that are available in centralized services. Once they have been implemented, smart contracts become immutable and are kept on every node of the blockchain network. This makes them resistant to manipulation since they cannot be altered. Some smart contracts have an upgrade capability, which allows the contract to be upgraded after it has been deployed \cite{16}. They enable the execution of a collection of commands along with access control, which helps to decrease the possibility of human negligence.

\section{Related Works}
\subsection{Creation And Transmission of Knowledge}
Creating new knowledge assets entails coming up with novel practical outcomes like new ideas, products, or concepts. Knowledge is developed via experience, communication, and instruction, wherein many forms of information are integrated and transformed \cite{17}. Esterhuizen et al.\cite{18} state that the process of creating new information has two aspects. According to their explanation, knowledge generation is viewed in a binary fashion by the first dimension. By this account, new bits of information are just like any other asset for an organization; they can be quantified and evaluated, and their value can rise and fall in response to changes in the external environment. The second point of view is the process dimension, which sees the process of knowledge generation as one that is not static but rather dynamic, interactive, process-oriented, and focused on relationships. Knowledge, according to Nonaka et al.\cite{19}, is formed through four processes: socialization, externalization, combination, and internalization. Finally, the authors conclude that the most productive environments for knowledge creation are those that encourage not just the process of innovation but also their evaluation and selection for use in resolving organizational challenges and capitalizing on economic possibilities.\\
The term "knowledge diffusion" was used by Appleyard et al. \cite{20} to describe the transmission and sharing of information and expertise between businesses. According to the authors, this spreading of information and ideas can take place through personal and professional networks. In \cite{21}, Wojick et al. emphasize the significance of the widespread transmission and utilization of information as a critical component in the growth of civil society. Knowledge transmission, as defined by Chen et al. \cite{22}, is the incorporation of previously acquired information into many fields of scientific and technical inquiry and advancement. Cohen et al. \cite{23} describe that the identification, accumulation of value, assimilation, transformation, and exploitation of knowledge resources to facilitate learning are all critical factors in the spread of information. In the context of this research, knowledge transmission refers to the process by which information is disseminated from its original point of origin to a wider audience by any means, including but not limited to word of mouth, formal education systems, and electronic media.
\subsection{Adoption of Blockchain Technology With Knowledge Marketplace}
The integration of blockchain technology and knowledge monetization results in a framework that is both more secure and easier to regulate when it comes to the exchange of information and services. This enables the company to develop a business model that is both scalable and flexible at a lower cost, which, in turn, leads to an improvement in the overall quality, reliability, and efficacy of research services. Cai et al. \cite{24} propose a novel, fully-fledged architecture for developing a trustworthy marketplace for knowledge based on streams of data gathered via crowdsourcing. Safe monetization of trustworthy insights gleaned in private from data streams for use in crowdsensing applications is supported by the proposed market architecture. This research makes use of lightweight cryptographic approaches to provide the first streaming truth discovery architecture for mining trustworthy information, and then it skillfully employs the blockchain to back up this monetization of knowledge while providing full security and service assurances.\\
In \cite{25}, Li et al. propose a framework for the cross-enterprise exchange of information and services in order to facilitate collaboration among manufacturers. The authors propose a distributed architecture for information and service sharing across manufacturing ecosystem participants, utilizing blockchain and edge computing as the enabling technologies. To avoid the issues of a centralized framework in which all information and knowledge are held by a limited number of parties, this one makes use of cutting-edge computing technology to create a flexible and dispersed network. The framework facilitates not just the sharing of information, but also the trading of expertise and services, allowing those involved to do their part. Standardization and protocols for its implementation are provided by blockchain technology, which also guarantees security and identity concerns thanks to the use of sophisticated data cryptographic algorithms.
\section{Operational Model of Knowledge Marketplace}
\begin{figure}[h] 
    \centering
    \begin{center}
    \includegraphics[scale=1]{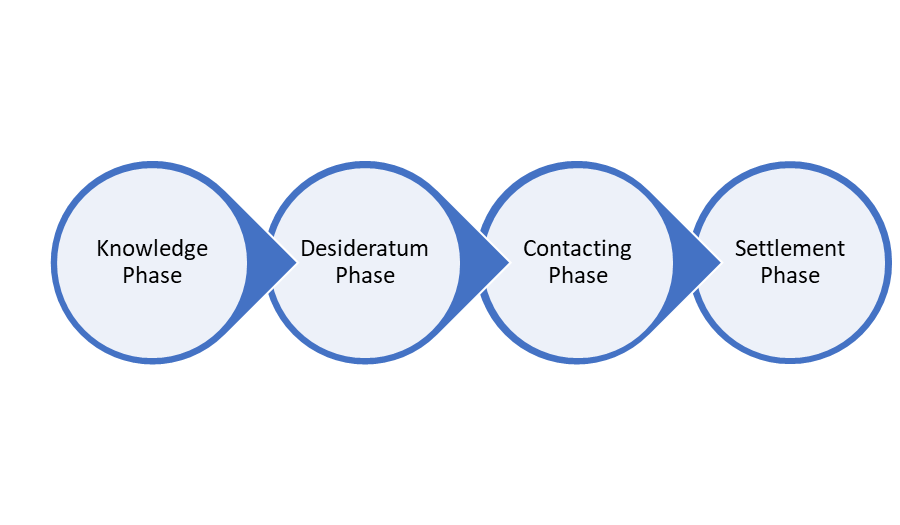}
    \end{center}
    \caption{Operational Structure Associated Knowledge Marketplace}
\end{figure}
In this section of the paper, the operational model of the knowledge monetization framework is investigated, along with its relevance in the overall process of information sharing. The functional model of The Knowledge Market is capable of being partitioned into a number of separate stages, which are enumerated in the following order:
\begin{itemize}
    \item \textbf{Knowledge Phase}: Providing market players with essential information about the items and services that are being offered is the primary focus of the knowledge phase. This information might be provided through a diverse method including but not limited to computerized marketing collaterals, services, or facilitators. When working with complicated goods like knowledge assets, the only way to ensure that the outputs of the knowledge phase will be satisfactory is to reach a consensus on the logical space. This consensus can take the shape of, for instance, a standardized lexicon with agreed-upon semantics.
    \item \textbf{Desideratum Phase}: During the desideratum phase, the market agent formulates specific plans to trade commodities and services with other parties. The end product has clear specifications for both the requests and the offerings. The computerized marketing collateral is the principal media that is used to make offers public to potential customers. The description of the offers has to be specific enough so that it may serve as a sufficient foundation for signing a contract.
    \item \textbf{Contracting Phase}: In the contracting phase, parties engage in negotiations, the successful conclusion of which is memorialized in a legally binding and cryptographically secure electronic contract. The results-oriented desideratum phase forms the basis of these contracts.
    \item \textbf{Settlement Phase}: The settlement phase necessitates services associated with finalizing the terms of the electronic contract finalized in the previous phase. Services such as monetization terms, knowledge ownership, information transfer, shipping, and insurance are all included here.
    \end{itemize}
   
\section{Conventional Knowledge Exchange System}
Existing methods of knowledge monetization are typically complex and have not undergone significant development in a very long time. Traditionally, the commercialization and exchange of knowledge takes place through the submission of possible research proposals, which then lead to research contracts. The entire process may be characterized as the following phases:
\begin{enumerate}
    \item \textbf{Research project stakeholders} \cite{26}: A considerable contribution to the successful completion of a project is made by its stakeholders \cite{27}. The following is a rundown of the numerous procedures that are involved in this phase:
        \begin{enumerate}
            \item Determine whom the project's stakeholders are by analyzing whether or not their interests and the project's interests are aligned.
            \item Determine the degree to which each stakeholder may influence the project.
            \item Put in place the necessary mechanisms to open up communication lines with the relevant stakeholders.

        \end{enumerate}
    \item \textbf{Developing a research proposal} \cite{28}: A template for research is a conventional paper that follows a preset framework and deviates from it only slightly. Several of these constituents may be broken down into the following categories \cite{29}:
        \begin{enumerate}
            \item Title 
            \item Introduction
            \item Literature Review 
            \item Methodology
            \item Plan - Timeline and schedule of activities
            \item Budget 
            \item Details of the research team (signed CV)
        \end{enumerate}
    \item \textbf{Identifying research funding}: The funding available dictates both the breadth of research topics that may be investigated and the nature of the findings that can be gleaned from that investigation. In most cases, financing is perceived in an overly simplified manner; there are national research councils, large charities, and government grants; nevertheless, other single funding organizations and instruments are being overlooked \cite{30}.
    \item \textbf{Costing and pricing of a research project}: Due to the fact that there are many different factors involved, the costing section of a study proposal is the most difficult element to complete. However, there are a number of other cases from which parallels may be made, and central research officials are also responsible for calculating the cost.\\
The most important distinction between estimating costs and setting prices is as follows:
        \begin{enumerate}
            \item The cost is figured by using the entire economic costing method (FEC). It takes into account all of the direct and indirect expenditures, including pay for employees, depreciation, space, central services, and a contribution to the university's infrastructure \cite{31}.
            \item Following the cost computation, the price is the amount that is requested.
        \end{enumerate}
    \item \textbf{Internal approval for research proposal}: There should be a consensus among all the entities involved in the research for submitting the proposal for research funding. This process should stand aligned with local policies before contacting any external funder.
    \item \textbf{Submitting a proposal}: The various entities should sign off on the research proposal before it is submitted. Most proposals are electronic, and various nuances accompanying this process should be considered and fulfilled. All the entities should register to the system for the proposal to be submitted.
\end{enumerate}
\section{Proposed Solution And Architecture}
The funding and financing system that exists is very opaque and has not evolved in a very long time. Modern solutions in the domain of blockchain can help evolve the system of funding so that both parties (researchers and funding organizations) can benefit from each other. Since the launch of the ETH mainnet in the year of 2015, there have been several open marketplaces, mainly for the purpose of selling tokens (mainly ERC721(NFTs)); the base architecture of these marketplaces can be used in constructing an open knowledge marketplace. The facets of knowledge monetization addressed by the proposed framework can be enumerated in the following order:
\begin{enumerate}
    \item Connecting researchers and funding associations - Various individuals or groups seeking funding can list their interest's in the form of listing in these markets. The funding associations (in the form of DAOs) can provide funding to these listings. Several or individual funding associations can together fund these listed research projects.
    \item Providing Data - Several individuals or organizations that wish to license data can do the same in knowledge marketplaces in a very efficient manner. Due to the transparent nature of blockchain-based marketplaces, data access can be checked. Due to this, the incentive gap can be filled.
    \item Building on top of the base layer -  Several data points can be tracked based on the interactions in the market. Due to the transparent nature, all the data generated in this market can be studied by individuals or organizations and advanced analytics can be generated from indexing the blockchain data, and statistics for researchers and funding associations will be widely available.
\end{enumerate}
\subsection{Connecting Researchers And Potential Funding Agencies}
\begin{figure}[h] 
    \centering
    \begin{center}
    \includegraphics[scale=0.85]{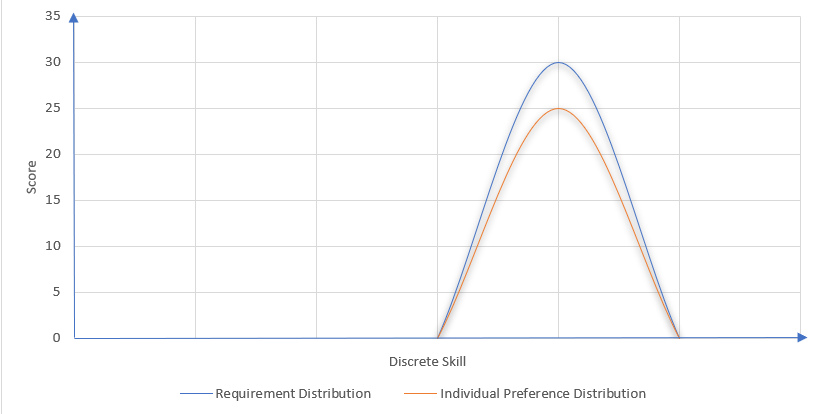}
    \end{center}
    \caption{Graph For Obtaining Cooperation Fit Index (Discrete Skill vs Score)}
\end{figure}
Active inference is a method that may be utilized for the purpose of assessing whether or not certain investors and researchers are compatible with one another. The design and upkeep of an autonomously functioning complicated system is one of the most important components of the Active Inference architecture. This may be utilized as a resource for the development of teams in research settings. The process involved in this mapping operation for synergistic collaboration can be summarized as:

\begin{enumerate}
  \item At time t = 0, the proposer presents a project and determines the distribution of skills needed for its completion. This distribution is then geometrically transposed onto a coordinate space with discrete skills as the axis, and the preferences of the Lab's members are then assigned to the corresponding radii. The area under the red curve (overlap) in Figure 3 is proportional to the cooperation fit index between the individual agents and the project.
  \item We determine the marginal probability of the research aim (o) and overlapping skill set (s) for participating researchers as shown in equation 1 at time t, when group strength has just crossed the threshold of 1, and use this information to determine the stability of the created system. 
  \begin{equation}
      \sum_{s} p(s,o)
  \end{equation}
  The proposed framework employs sophisticated logic to discover a solution because estimating marginal probabilities is frequently computationally difficult (a corollary to surprise minimization in Active Inference theory, equation 2).
   \begin{equation}
      F = \sum_{s} q(s)\ln\frac{q(s)}{p(s,o)}
  \end{equation}
  Creating an upper bound by using the identity of the Jensen inequality (equation 3), which is termed as Free energy (F). Now, minimizing the Free Energy is the optimization challenge for the stability of the system.
This procedure changes states repeatedly (like adding or removal of a member).
 \begin{equation}
      -\ln \sum_{s} q(s)\left[\frac{p(s,o)}{q(s)}\right]\leq -\sum_{s}q(s)\left[\ln\frac{p(s,o)}{q(s)}\right] = F
  \end{equation}
  \item The proposed framework must now create sub-groups in order to have a functional research niche. This might be accomplished by algorithmic permutation throughout the range of potential configurations, from completely symmetrical sub-groups (each having the same number of members) to perfectly asymmetrical groupings (each with different numbers). In Algorithm 1, the quantized stability quotient for simultaneous creation and existence is shown. Although this "selection of project groups" is carried out by Humans (through Affordances, in Roles), it may be viewed as a choice of organizational policy. The set of potential teams in this situation (policy evaluation at lab size over team composition) is made up of all conceivable deployment potentials of eligible participants into project-specific roles. Given that this affordance space is discrete and finite, it might be able to precisely design and carry out this group formation strategy choice.

\end{enumerate}
\begin{algorithm}
\caption{Finding Stable Sub-Group Configuration}
\begin{algorithmic}[1]
    \State$F \gets Free\ Energy$
        \State$n \gets total\ system\ strength$
        \State$sn \gets sub-group\ strength$
        \State$g \gets possible\ sub-group\ strength$  \Comment{range 2 to n; all configurations}\\
    \For {$\texttt{i in (g/sn)}$}
        \State \texttt{F[i]=sum(F(g,sn))}   \Comment{Perfectly Symmetrical Sub-Groups}
    \EndFor
    \For{\texttt{i in (g[i]/sn[i])}}
        \State \texttt{F[i]=sum(F(g,sn))}   \Comment{Perfectly Asymmetrical Sub-Groups}
    \EndFor
    \end{algorithmic}
\end{algorithm}
\subsection{Monetization And Exchange of Knowledge Assets}
\begin{figure}[H] 
    \centering
    \begin{center}
    \includegraphics[width=\textwidth]{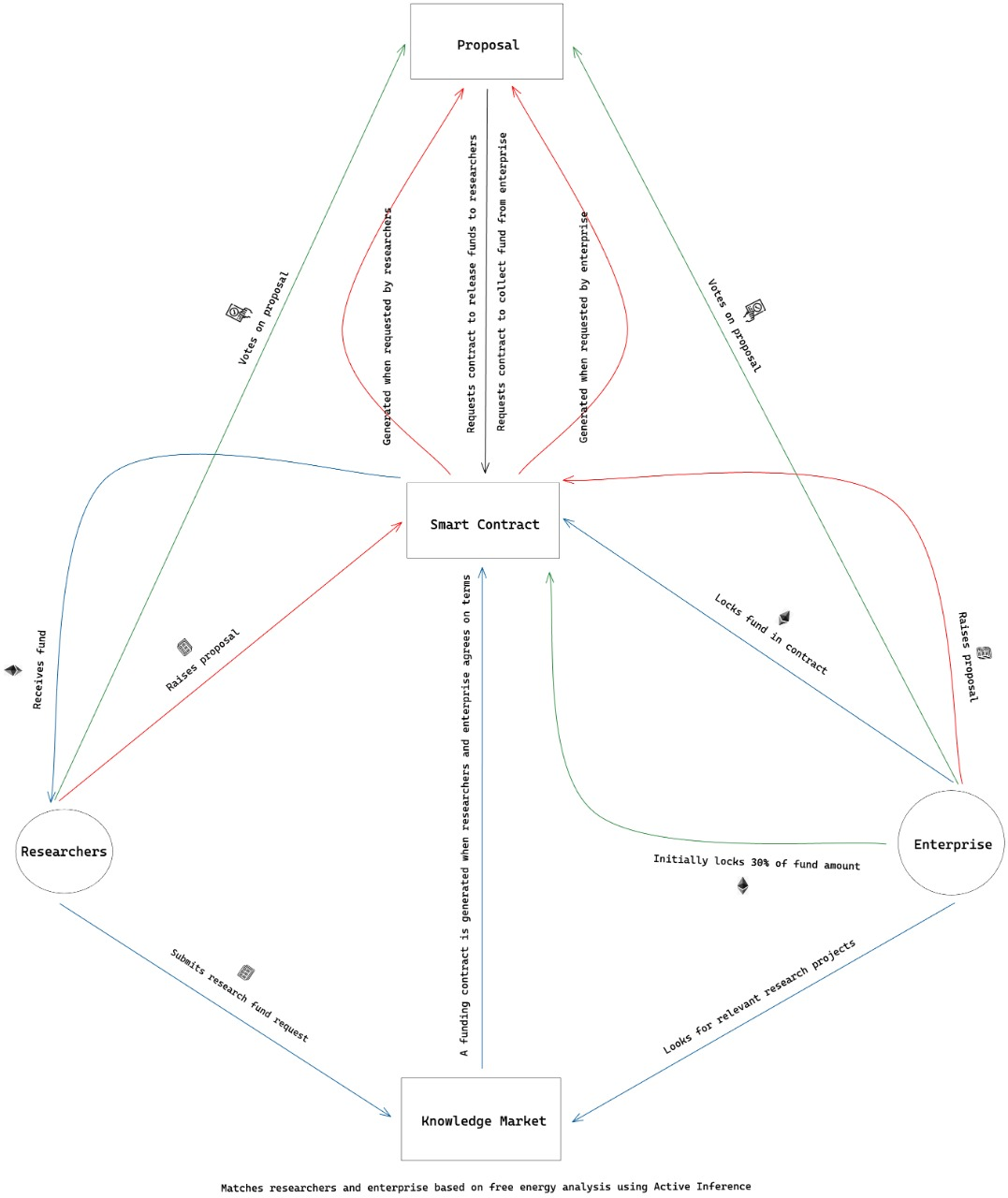}
    \end{center}
    \caption{Roles And Processes Associated With The Proposed Framework}
\end{figure}
Within the scope of this portion of the paper, the authors investigate the method of monetizing knowledge assets and exchanging them, which is illustrated in Figure 4. Both of these entities, investors and researchers will be brought together through the formation of knowledge markets; once an association has been established, the steps outlined below can be taken:
\begin{enumerate}
    \item Using a known structure in traditional decentralized funding via the means of DAO: One of these traditional funding schemes is called DAICO (DAO + ICO). DAICO was proposed by Vitalik Buterin in 2018 to make the process of ICO more secure for investors during the initial stages of the development process. The investors have control over the funds, and part-wise allocation based on the proposal submitted by the seekers can be voted on, and based on the votes, payouts can be made.
    \item This approach helps form the following affordances:
    \begin{enumerate}
        \item The researchers will raise a proposal for the undertaking in a standardized template.
\item The investors will go through the proposal and will determine the value that the undertaking brings to the entirety of the project (actuarial analysis). 
\item After the analysis, the investors will vote on the proposal; if 51\% of the votes are in favor of the proposal, then the funds get released to the researchers. In the case that the proposal gets less than 51\% percent of the votes, the investors will have to provide the reason for the rejection of the proposal. Researchers can redo the proposal and resubmit it with modifications in accordance with the comments provided by the investors.
\item In case there is a prolonged disagreement between the two parties, there can be a proposal for forfeiting the bond. This disagreement can be determined mathematically with the use of free energy analysis.

    \end{enumerate}
    \item The funders will have to provide about 30\% of the promised amount up front in an escrow account, this may be released on completing milestones, but this will be the minimum guaranteed amount. In case the funders don't plan to pursue the whole research endeavor, the researchers will be guaranteed the 30
\end{enumerate}
\section{Conclusion}
In this paper, the focal objective was to investigate the feasibility of integrating blockchain technology and active inference with the knowledge monetizing sector. The paper proposes a decentralized framework to function as a knowledge and services exchange platform by embedding the recent technologies in blockchain and active inference. To ensure the knowledge is fairly and securely monetized, authors have proposed a comprehensive design that makes use of blockchain technology, which is becoming increasingly popular, to ensure the knowledge is monetized with the guarantees of monetary impartiality, knowledge confidentiality, and refined automating. Moreover, the proposed framework also employs active inference that ensures a synergistic collaboration between the stakeholders of the knowledge marketplace.
The authors plan to advance toward the notion of digital ownership and further develop the offered solution of mapping the potential investors and researchers across the network. The authors want to evaluate their suggested method across many blockchain platforms and consensus mechanisms.

\bibliographystyle{unsrt}  
\bibliography{references}  
\end{document}